\documentclass{article}

\usepackage{geometry}                
\usepackage{graphicx}
\usepackage{amssymb}
\usepackage{epstopdf}
\usepackage{slashed}
\usepackage{hf-tikz}
\usepackage{footnote}
\makesavenoteenv{eqnarray}
\usepackage{amsmath}

\makeatletter
\newcommand{\manuallabel}[2]{\def\@currentlabel{#2}\label{#1}}
\makeatother

\manuallabel{ch:N-1}{6}
\manuallabel{ch:N-2}{7}
\manuallabel{ch:N-3}{8}
\manuallabel{ch:N-4}{9}

\manuallabel{sec:stringphys}{6.2}
\manuallabel{sec:StringQ}{6.3}

\manuallabel{fig:scat}{8.1}

\manuallabel{eq:closed}{6.6}
\manuallabel{eq:Ssig}{6.16}
\manuallabel{eq:ham}{6.20}
\manuallabel{eq:HamMass}{6.36}
\manuallabel{eq:open}{6.9}
\manuallabel{eq:sigmin}{6.21}

\usepackage{natbib}
\bibpunct{(}{)}{;}{a}{}{,}

\begin{document}

\title{Chapter \ref{ch:N-2}: Duality}

\author{Nick Huggett and Christian W\"uthrich\thanks{This is a chapter of the planned monograph \emph{Out of Nowhere: The Emergence of Spacetime in Quantum Theories of Gravity}, co-authored by Nick Huggett and Christian W\"uthrich and under contract with Oxford University Press. More information at www.beyondspacetime.net. The primary author of this chapter is Nick Huggett (huggett@uic.edu). This work was supported financially by the ACLS and the John Templeton Foundation (the views expressed are those of the authors not necessarily those of the sponsors). We thank Dave Baker, Neil Dewar, Doreen Fraser, Brian Greene, Jeff Harvey, Keizo Matsubara, Joshua Norton, James Read, Tiziana Vistarini, Eric Zaslow, and two anonymous referees for help at various stages of this work. This chapter is a revised, updated, and expanded version of \cite{Hug:15}.}}

\maketitle

\tableofcontents

\ 

In this chapter we will see how string theory contains some surprising symmetries -- `dualities' -- which, we will argue, put pressure on the view that the spacetime in which strings are described can be literally identified with classical, physical spacetime -- instead it is `emergent' from the theory. While the following stands on the previous chapter, and exemplifies its physics, it can be read on its own  to understand the essential conclusions. We focus on one such symmetry, `T-duality', but at the end review others.

\section{T-duality}

Consider a closed, classical bosonic string in Minkowski spacetime with a compact spatial dimension, $x$, of radius $R$.\footnote{Our technical presentation follows, amongst others, \cite{BraVaf:89}, \citet[237]{Gre:99}, \citet[Ch 17]{Zwi:04}, and \cite{Zas:08}. The treatment given here uses an approximation based on the `double field theory' approach (see e.g., \cite{AldMar:13}).} As we saw in the previous chapter, its state is a function 
$X(\sigma,\tau)$ describing the $x$-coordinate of the point of the string worldsheet with worldsheet coordinates $(\sigma,\tau)$: hence the state of a string specifies an embedding of the worldsheet into spacetime. 

\begin{figure}[htbp] 
   \centering
   \includegraphics[width=2in]{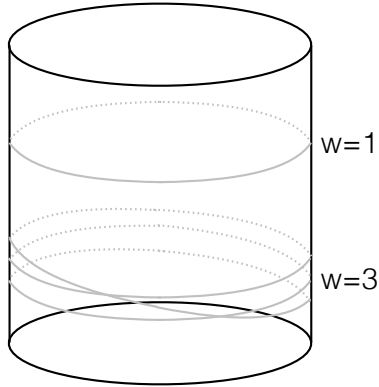} 
   \caption{Two closed strings in a 2-dimensional space with one compact dimension. One string is wrapped once around $x$, and the other three times -- with winding numbers $w=1$ and $w=3$, respectively.}
   \label{fig:wound}
\end{figure}

We adopt the conventions that the spacelike string coordinate $\sigma$ is periodic with period $\pi$, while the compact spatial coordinate $x$ is periodic with period $2\pi$ (so that $\sigma$ and $\sigma+n\pi$ are labels for the same point of the string worldsheet, and $x$ and $x+2n\pi$ are labels for the same point of space). Then the state of the string with respect to $x$ is

\begin{equation}
\label{eq:wound}
X(\sigma,\tau) = 2w\sigma R + 2\ell_s^2p\tau + \mathrm{vibrational\ terms}.
\end{equation}
This expression differs from equation (\ref{eq:closed}) by the addition of the first term, which describes the $w$-fold winding of the string: for instance, if the string is wound once around $x$, so $w=1$, then $X$ ranges from 0 to $2\pi$ as $\sigma$ ranges from 0 to $\pi$. The second term represents the linear momentum of the string; the constant $\ell_s$ is the `characteristic string length'. For simplicity, we shall ignore vibrations, since they do not change the substance of this chapter.\footnote{\label{ftnt:rescale}Important: to clean up expressions in this chapter, we rescale $\ell_S$ by a factor of $1/\sqrt2$, and, in (\ref{eq:HamP}), $T$ by a factor of $\pi/4$, relative to chapter \ref{ch:N-1}. Though there is a small risk of confusion arising, we believe this decision makes this chapter more accessible to those reading it without studying the previous chapter. These unconventional changes have no physical significance; indeed, one could have achieved most of the same result by trivially rescaling the worldsheet coordinate.}

In wisely chosen string coordinates (and suitable units), substituting $X$ in the Hamiltonian (\ref{eq:ham}) gives 

\begin{equation}
\label{eq:HamP}
H =  \frac{T}{8\pi}\int  \dot{X}^2 + X'^2\ d\sigma =  \frac{T}{2\pi}\int (\ell_s^2p)^2 + (wR)^2\  d\sigma,
\end{equation}
where $T$ is another constant, the string tension. Not surprisingly there is a kinetic term, plus a term from the winding, hence stretching, of the string around the closed dimension. The next step is to quantize.

Momentum first. The closed dimension implies a periodic boundary condition for momentum eigenstates $\Psi_k(x)=e^{ikx}$ (ignoring normalization, and with $\hbar=1$)

\begin{equation}
\label{ }
\Psi_k(0) = \Psi_k(2\pi R) \Rightarrow e^{ik\cdot 0} = 1 = e^{ik\cdot2\pi R} \Rightarrow k=0, \pm1/R, \pm 2/R\dots.
\end{equation}
In other words, momentum is quantized: $|k|=n/R$, with `wave number' $n$. Substituting into the Hamiltonian (\ref{eq:HamP}), we obtain the spectrum

\begin{eqnarray}
\label{eq:Hspec}
E_{n,w} & = & \frac{T}{2\pi}\int (\ell_s^2n/R)^2 + (wR)^2\  d\sigma.
\end{eqnarray}

Now winding. Assuming interactions, in QM a string can change the number of times it is wound around a closed dimension. Hence $w$ is not a constant, classical `c-number' of the system, but  a dynamical \emph{quantum} quantity, described by a wavefunction. Pay close attention to this point, as it is \emph{crucial}:  because the winding number can change over time, a quantum string can be in superposition of states of different winding numbers. Without this move there is no T-duality -- in this sense it is a quantum phenomenon.

The winding term in (\ref{eq:HamP}) depends on $l=wR$, which must have a discrete spectrum since $w$ does. Thus much as before, these eigenstates have the form $\Phi_l(y)=e^{ily}$ around a circle with coordinate $y$, but with radius $1/R$.\footnote{This quantity seems to have units of $length^{-1}$, but the numerator can be taken as an \emph{area} to give overall correct units. A similar point applies everywhere that quantities appear to have the wrong units.} In that case the periodic boundary condition $\Phi_l(0)=\Phi_l(2\pi/R)$ yields

\begin{equation}
\label{ }
e^{il\cdot0} = 1 = e^{il\cdot2\pi/R} \Rightarrow  l= 0, \pm R, \pm2R, \dots = wR,
\end{equation}
as required. Overall then, the state of a quantum string involves (the tensor product of) two wavefunctions, one representing its position/momentum, and another representing its winding.

The question is of course, `where is the circle on which the $\Phi_l(y)$ wavefunction lives?' It can't just be in space, because then $\Phi_l(y)$  describes a second object which we could expect to find somewhere. Instead, there must be a new `internal' dimension associated with each closed dimension of space; hence Witten calls $y$  `another ``direction'' peculiar to string theory' \cite[29]{wit96}. His proposal is not that \emph{space} has an extra dimension for every dimension a string can wrap around, but rather that treating winding as a quantum observable means that it can be represented like momentum on a \emph{non-spatial} circle. Or more precisely, when we consider the space of all states of any momentum or winding, we find two quantum `position' operators, $x$ and $y$, respectively corresponding to position in physical space (radius $R$) and in a new `winding space' (radius $1/R$). But observables represent physical quantities, so we have to take both `positions' and spaces equally seriously, even if only one is physical space; let's call the other `winding space'. But remember, the string winds around physical space, while the winding number wave lives in winding space. (Take the term `physical space' with a grain of salt here: it is the space in which the string moves, but precisely because of T-duality we will have to clarify below exactly how it relates to ordinary, observable, classical space.)

\ 

\setlength{\leftskip}{1cm}

\noindent \emph{Semi-technical aside}: as usual, $x$ and $y$ are `position' operators for physical and winding space. Moreover, as $\hat p=-i\partial/\partial x$ is the momentum observable with eigenvalues $k=n/R$ in the periodic plane wave states $e^{inx/R}$, so $\hat w=-i\partial/\partial y$ is the winding observable with eigenvalues $l=wR$ in the winding states $e^{iwRy}$. Thus each space is associated with identical canonical commutation relations, $[\hat x,\hat p]=i$ and $[\hat y,\hat w]=i$ (the observables from different spaces commute). Therefore, since position and momentum generate the algebra of observables, each space has, formally speaking, exactly the same observables, individuated as functions of $x$ and $\hat p$ or $y$ and $\hat w$.\footnote{Of course there are observables involving operators from both the spaces, but since the latter commute, such observables are always the commutative product of a pair of observables, one from each space. So all the points we need go through trivially, and we will ignore them.}

\setlength{\leftskip}{0cm}

\ 

\begin{figure}[htbp] 
   \centering
   \includegraphics[width=4in]{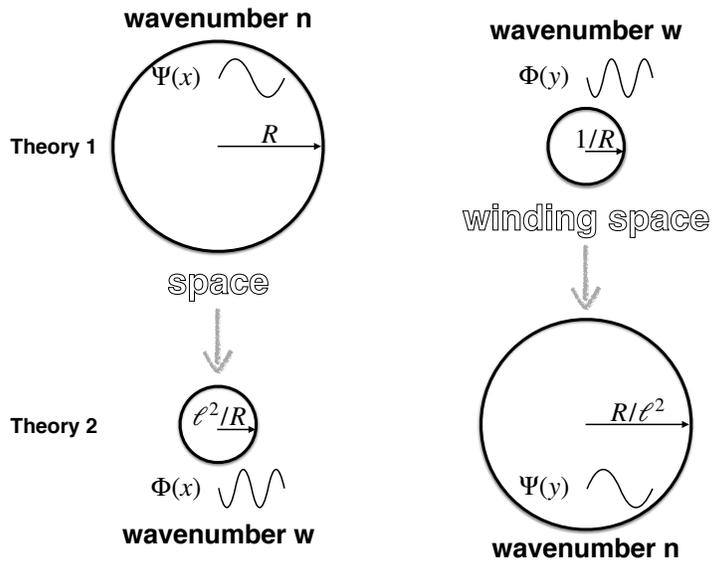} 
   \caption[Caption for LOF]{Reading horizontally first, Theory 1 describes a string moving in a space $x$ of radius $R$ with wavenumber $n$ -- more generally, with wavefunction $\Psi(x)$; and winding described by the wave number $w$ -- more generally, wavefunction $\Phi(y)$ -- in a `winding space' $y$ with reciprocal radius, $1/R$. Now reading vertically, a second, `dual' theory is obtained by simultaneously taking $R\to\ell^2/R$ and $n\leftrightarrow w$ (equivalently, $\Psi\leftrightarrow\Phi$).\footnotemark}
   \label{fig:tdual}
\end{figure}

Such `internal' spaces are familiar -- for spin states and gauge field states, for instance -- so there is nothing new yet. But look again at (\ref{eq:Hspec}), the spectrum of the Hamiltonian. It's easy to check that a string with wave number $n$ and winding number $w$ in a space with radius $R$ has the same energy eigenvalue as a string with \emph{winding} number $n$ and \emph{wave} number $w$, \emph{but which lives in a space of radius} $\ell_s^2/R$: 

\begin{equation}
\label{eqn:T-D}
n \leftrightarrow w \quad \mathrm{and} \quad R \to  \ell_s^2/R.
\end{equation}
The  second string has a spatial wavefunction in a compact dimension of radius $\ell_s^2/R$, and hence -- by the same reasoning as before -- a winding wavefunction that lives in a compact dimension of reciprocal radius, namely $R/\ell_s^2$. If the first string lives in a space with radius $R>\ell_s$, then the second string lives in a space of radius $\ell_s^2/R<\ell_s$: the strings are `reflected' through $\ell_s$.  See figure \ref{fig:tdual}. 

\footnotetext{In the next section we will see that the label `space' should be replaced with `target space'.}

Now, $n$ and $w$ label eigenstates of momentum and winding, and so in terms of wavefunctions,  $n\leftrightarrow w$ corresponds to $\Psi_n(x)\to\Phi_w(x)$ and $\Phi_w(y)\to\Psi_n(y)$: the wavefunctions are exchanged between space and winding space. Thus for \emph{general} spatial and winding wavefunctions (i.e., superpositions of momentum or winding eigenstates) $\Psi(x)$ and $\Phi(y)$, respectively, let $\Psi(x)\to\Phi(x)$ and $\Phi(y)\to\Psi(y)$: the `dual' string has the same -- but exchanged -- wavefunctions. (And if we took account of the vibrations of the string, these would also be exchanged between physical and winding spaces.)

Then, ($i$) because they have the same Hamiltonian, both strings will have the same mass spectrum (because in string theory the Hamiltonian determines the mass: see equation \ref{eq:HamMass}). Moreover, ($ii$) because the roles of momentum and winding are reversed in the Hamiltonian by (\ref{eqn:T-D}), the dynamics of the spatial wavefunction in one string become the dynamics of the winding wavefunction in the other, and vice versa: in other words, the strings remain dual over time under the exchange of momentum and winding.

Further, ($iii$) because momentum and winding are exchanged by (\ref{eqn:T-D}), every observable pertaining to physical space is exchanged with a corresponding observable pertaining to winding space; and because the wavefunctions are also exchanged, the expectation value of the new observable for the dual string will equal that of the original observable for the original string. (And vice versa.) In short, the \emph{pattern} of observable quantities will be preserved by (\ref{eqn:T-D}); what changes is whether the quantity is understood to pertain to physical or to winding space.\footnote{The mapping introduces a $\ell_s^2$ factor, but these  can be absorbed in a trivial rescaling of observables, so we will ignore it.} (And similarly once one includes the vibrations of the string.)

\ 
\setlength{\leftskip}{1cm}

\noindent\emph{Continuation of the semi-technical aside}: a little more formally, the point is that the algebra of observables on spatial wavefunctions for one string is mapped onto the identical (as we saw above) algebra of observables on winding wavefunctions of the other -- with $x\leftrightarrow y$ and $\hat p\leftrightarrow\hat w$. Since the wavefunctions are also exchanged, the values associated with \emph{all} corresponding elements of the algebra of observables are preserved by (\ref{eqn:T-D}) -- the entire pattern of expectation values. (But generally not the expectation values of  specific operators.)

\setlength{\leftskip}{0cm}

\ 

Thus the systems are equivalent in the following sense: the Hamiltonian and hence dynamics are the same, and the pattern of physical quantities (formally represented by observables) agree. This equivalence, and others comparably strong, are known as `dualities', and (as implicitly anticipated) the two theories related by it are `dual' to each other, or `duals'.\footnote{There are competing accounts of what exactly makes a symmetry a `duality' (or even a `symmetry'). We have in mind an account along the lines of that given in \cite{Har:19}, which aligns with our interpretative goals. This sense may not align perfectly with traditional use of the term, nor (according to \cite{Daw:17}) its novel methodological meaning for string theorists. See also \cite{De-:19}.} In particular, (\ref{eqn:T-D}) is known as `T-duality', where -- depending on whom you ask -- `T' either stands for `target space' (i.e., the space in which the string lives), or for `torus'. T-duality holds not only for the bosonic string just considered, but also for supersymmetric string theory, in which case it also changes the character of the string: for instance, type IIA strings are T-dual to type IIB strings.

There is a lot to unpack and justify in these statements, which will be the work of the following section, but for now a concrete example, taken from \citet{BraVaf:89}, will help illustrate some of its implications. Take T-duals, $T_1$ and $T_2$, which differ in the radius that they postulate for a closed spatial dimension: a circumference of $10^{12}$ light years (two orders of magnitude bigger than the visible universe) on the one hand, and $10^{-94}$m on the other (assuming a value for the characteristic string length of $10^{-33}$m, two orders of magnitude above the Planck length). Thus $T_1$ and $T_2$ (apparently) make radically different assertions about the size of a spatial dimension. Before T-duality, one would assume that simple observations would rather readily choose between them, but that can't be right if the duals are physically equivalent.

To understand how the equivalence manifests itself, in a beautiful conceptual analysis, Brandenberger and Vafa consider an archetypical measurement of the radius: fire off a particle of known velocity -- a photon, say -- and time its journey around space. Suppose the result is a trillion years: that seems pretty conclusive evidence for the large radius story, $T_1$. But, in terms of $T_1$, how is the measurement described? The photon has a spatial wavefunction $\psi(x,t)$, which evolves, according to the Hamiltonian, from being localized nearby, via a journey away of $10^{12}$ years, to being localized nearby again. However, $T_2$ can also account for this result. 

Note that the photon is a low energy mode of the string, the easiest thing to excite. Indeed, using the Planck tension of about $10^{44}$N as an estimate of the string tension, we find that in $T_1$ the first excited winding state corresponds to the mass-energy of $10^{15}$ supermassive black holes; so `low' energy here is relative to almost inconceivable scales! On the other other hand $k=n/R$, so momentum is inverse to the radius of space, and the very large radius of space in $T_1$ allows states of very low momentum, hence of low energy. In other words, in $T_1$ the low (by any sensible measure) energy photon used to measure the circumference has a state involving \emph{only} momentum excitation, and the lowest, $w=0$, winding state.

But in $T_2$, with reciprocal spatial radius, even the smallest excitation has tiny wavelength and sohuge momentum, while the energy of stretching a string around a dimension of radius $10^{-94}$m is tiny (about a tenth of the electron mass). Thus in $T_2$ the (same) low energy photon in the experiment is described by winding modes, and the lowest, $k=0$ momentum state. And of course the states are indeed dual: a $T_1$ momentum state maps to a $T_2$ winding state, the former with wave number, and the latter with winding number, $n$. Thus, according to $T_2$ the photon has the dual state represented by the same wavefunction $\psi(y,t)$ \emph{but in winding space}. Then, because the Hamiltonian is the same, but with the roles of physical and winding space reversed, the photon evolves in exactly the same way -- namely a `journey' around \emph{winding} space, taking $10^{12}$ years, as observed. No surprise there: in $T_2$ physical space is tiny, hence winding space, with reciprocal radius is huge. (Of course, from the point of view of physical space, the evolving winding wavefunction describes some changing superposition of states in which the string is wound different numbers of times around the tiny radius.)

This analysis shows that because the experiment is characterized as timing a low energy particle (of a given type), by construction it involves a process in the larger of physical and winding space, and so is guaranteed to take a long time, guaranteed to produce the phenomena of a large radius space in either dual. (Indeed, because they are dual, the process is guaranteed to take the same time, be an observation of the same large radius.)

And the equivalence generalizes. Any process in \emph{physical and winding} space according to $T_1$, corresponds to a process in $T_2$ in \emph{winding and physical} space, and so no measurements or observations of even the most hypothetical kind will distinguish them. And so, the question comes up of which we should take to be correct; or indeed, whether the difference between tiny and huge is a true physical one at all. In the next section we will review some responses to this situation, and give reasons to favor one.

At this point we want to acknowledge that  assuming that physical states supervene on expectation values is to take a strong stance on the interpretation of QM. For instance, that assumption is clearly false according to Bohm's theory. Moreover, \cite{Nik:07} has shown that Bohmian string theory breaks T-duality as a symmetry. While we take the Bohmian view very seriously, in this discussion we will explore the consequences of duality for interpretations in which there are no `hidden' variables.

\section{T-duality and the nature of space in string theory} 

Having explained T-duality in the previous section, we now turn to unpacking its significance for the nature of space in string theory (and by extension, the possible significance of other dualities). In \S\ref{subsec:tax} we will lay out different ways in which one might take a duality. We will consider two interpretational forks: first, whether, despite appearances, duals are physically equivalent, in fact representing the same physical states of affairs; or whether they are inequivalent, but equally compatible with any observation. Second, we will see that the way that the duals represent makes a difference to what the state of affairs on which they agree is -- what, that is, they say about spacetime.\footnote{Note that \cite{Mat:13}, \cite{Read:2014fk}, and \cite{Le-Rea:18} provide alternative complementary categorizations of the stances on duality, useful in the context of different  background philosophical issues. \citet{DieDon:15} proposes a distinction similar to my second taxonomic fork.}  Then we turn to deciding which interpretational choices are best: we argue in \S\ref{subsec:D=PE} for the physical equivalence of the duals, and in \S\ref{subsec:fact?} that the duals describe strings living in a space of no determinate radius -- the main conclusion of this chapter.

\subsection{Interpretive questions}\label{subsec:tax}

T-duality provides a mapping (\ref{eqn:T-D}) between a pair of theories that agree (under the mapping) on the expectation values of all observables in all states, and on the evolutions of all states. 

Note that there is a mismatch between the philosophers' and the physicists' use of `theory' at this point. Roughly, physicists distinguish theories by the mathematical form of the laws (or action), while philosophers often further distinguish them by the values of their constants. In physicists' but not philosophers' terms, T-duals are the same theory, but there is no substantive disagreement about the facts. We have decided to follow philosophers' sense of `theory' here, because it will facilitate the following discussion to emphasize the differences between the duals. We however follow the physicists in describing T-duality, and other dualities in which the laws take the same form in the duals, as `self-dualities'. (Later we will describe `holographic duality', which is not a self-duality.)

It's also important to emphasize that `observables' here does not have any narrow philosophical empiricist meaning: it denotes the collection of hermitian operators (subject to any selection rules), \emph{not} some `special' collection of properties to which we have especially `direct' access. Indeed, the observables are thus those operators normally thought of as representing the totality of physical, quantum mechanical quantities, including those far from immediate `observation' by humans. And with respect to those quantities the theories are -- under the mapping -- in perfect agreement.

The qualifier `under the mapping' is crucial. \textit{Prima facie}, in one system a string has momentum $n/R$, and is wound $w$ times around a dimension of radius $R$. In the other, it has momentum $Rw/\ell_s^2$ and is wound $n$ times around a dimension of radius $R'\equiv\ell_s^2/R$. And in the quantum mechanical treatment spatial and winding-spatial parts of the wavefunction are interchanged: $\Psi(x)\otimes\Phi(y)\to\Phi(x)\otimes\Psi(y)$  (which, in the case of simultaneous momentum and winding \emph{eigen}states, entails $n \leftrightarrow w$). If the physical interpretation of the operators is held fixed, then the theories are inequivalent. So the crucial question will be that of their interpretation.

Normally also one thinks of c-numbers such as $R$ as physical, in which case duals again describe different physical situations. But normally, c-number parameters can be determined by the values of quantum quantities: the charge on the electron, say, by scattering probabilities. A duality arguably means they cannot be so determined by the values of the observables of the theory: the pattern of expectation values is preserved. So we should at least leave open that such differences in the c-numbers do not, after all, represent physical differences. In \S\ref{subsec:D=PE}, invoking a simple duality, we will argue that indeed they do not. But for now we have our \emph{first interpretive decision}: either the T-duals agree on the physical world or they do not. If they do, then for the purposes of this enquiry they say the same, \emph{all in}; we will not be interested in any putative non-physical differences. 

Most commentators have agreed that the T-duals should indeed be taken as giving the same physical description: especially, see \cite{Dawid:2007gf}, \cite{Mat:13}, \cite{Rickles:2011db}, \cite{Rickles:2013ul}, and \citet[chapter 3]{Vis:19}.  However, there have been recent skeptical or dissenting voices: especially, \cite{ReaMol:18}, \cite{Wea:19}, and \cite{But:19}. We will argue below (\S\ref{subsec:D=PE}) that physical equivalence is a reasonable conclusion, but we will also clarify how this should be understood, and modify the position taken in \cite{Hug:15}, in the light of the more recent work. String theorists often seem to endorse physical equivalence of duals (for instance, \citealt[247]{Gre:99}). However  their words sometimes seem ambiguous on the point. For instance, \cite{Teh:2013rc} identifies remarks suggesting that one dual may be more fundamental than another\footnote{See also Read and M\o ller-Nielsen. However, we will discuss in \S\ref{sec:GGD} a sense in which relative fundamentality does play a role in holographic duality.}

, and an important point made by Harlow and reported at the end of this chapter).  However, taking the view that duals are physically equivalent, a second interpretive decision awaits.\footnote{We want to thank Dave Baker for emphasizing that there are distinct options here.}

To describe the options now facing us, it is first necessary to be more careful in distinguishing the different conceptions of space that have entered the discussion. So far we have already distinguished `physical' and `winding' spaces, but to proceed we need to revise this division. First, the former concept will be bifurcated. Second, it should be clear by now that winding space is every bit as `physical' as `physical space', so we will drop that terminology as misleading. 


Instead we start from an analytical distinction between `theory' and `phenomena'. That is, when a new, more fundamental scientific theory explains an established, less fundamental theory, which has stood the test of experiment, then we can speak of the latter as the `phenomena' \emph{relative} to the former. Being relative, the distinction is suited for the historical process by which today's novel `theory' becomes experimentally vindicated, and eventually becomes tomorrow's bedrock empirical given: Kepler's laws were phenomena for Newton's, but the laws Newton inferred from them were themselves phenomena for general relativity.\footnote{It is not essential for the reader to accept this historical picture. It's helpful to accept a theory-phenomenon distinction, which has its origins in \citet[1-19]{Car:83} and \citet{BogWoo:88}. What we take from them is the idea that `phenomena' are abstracted from direct observation events, and so have a `theoretical' structure themselves: We then add that a theory might therefore be phenomenal relative to a more fundamental theory.} In the present case, the more fundamental theory is string theory, which aims to explain, amongst other things, our current account of space. This account is expressed in our current best scientific theories, quantum field theory and general relativity, in the small (high energy physics) and large (cosmology). All of these are the `phenomena' relative to string theory. In \cite{Hug:15} the space that they describe was thus called `phenomenal'; however, this term has suggested to some the mental content of spatial experience rather than a physical structure, so in this book we have used the terms `classical', or `relativistic', or  `observed', or plain `space' instead. Under this concept we also include the  looser prescientific geometrical space we take ourselves to observe in the everyday, including the experience of three large dimensions. (Of course, quantum field theory, general relativity, and everyday experience describe space in strictly incompatible ways: flat and curved, relativistic and not. But the relations between these descriptions, especially as limits of each other, are well-enough understood to make the notion of a single classical space, described by these phenomenal theories, clear enough for our purposes.)

In contrast, any `space' that appears in the formulation of string theory is `theoretical' in the sense described. There are in fact two theoretical spaces in (quantum) string theory: the first, of course, is winding space. The second is the space in which strings (and their momentum wavefunctions) live, which in the previous section we equated with classical space, under the concept of `physical space'. But, according to the theory-phenomena distinction, classical space is a `phenomenal' physical space, while strings live in a `theoretical' physical space. The ultimate question of this section is of the relation between these two spaces, whether they are identical, or whether one is reducible in some sense to the other. To clarify that investigation we thus adopt a new term -- `\emph{target space}' -- for the latter, the space in which strings propagate. (There are two reasons not to use `physical space' for this theoretical notion: first, as noted all three of the concepts of space we have discussed are `physical' in a general sense; second, on one interpretation of T-duality, target space will turn out to have novel features, and the novel name will avoid the inapt connotations that would come with a pre-existing concept.) `Target space' is a term of art in string theory, referring to the `background' space in which a string is embedded: by the function $X(\sigma,\tau)$. The classical string is literally located in target space, and wound around it; we will take it that it makes sense to extend this intuitive picture to the quantum string, which thus also `lives' in a background target space. As we have formulated the theory in this chapter, this situation is represented by the string's position/momentum wavefunction being a function over target space: just as we represent a quantum particle being `in' a region by a non-zero wavefunction over that region. It is natural when first introduced to string theory, to think that target space is simply the same space we ordinarily experience, or at least space as conceived in contemporary physics: T-duality makes this identification problematic. Hence figure \ref{fig:tdual} should be modified, with `space' replaced by `target space'.

Given that general relativity and quantum field theory (and our everyday understanding) are the context of phenomenal space, measurements of its radius are operationalized in their terms: as in Brandenberger and Vafa's thought experiment, for example, which appeals to the photons and clocks of extant physics. Thus the radius of classical space, as defined, is given by $c$ times the duration of the photon's journey. In terms of such measurements, classical space is observed to be very large: we don't know its radius (or even whether it is compact), but we can observe $10^{10}$ light years of it -- and even a simple glance around the room shows that it is much larger than $10^{-33}$m!\footnote{In fact we will count any additional `small', dimensions also as `classical': though they may be required by certain theories of quantum gravity, their possibility is not at all quantum mechanical, as the original Kaluza-Klein theories demonstrate. Even though they are microscopic relative to the ordinary dimensions, they may still have a large radius on the relevant scale, $R>\ell_s$.} Moreover, we have also seen how the dual theories will both predict that empirical result. While giving dual descriptions of the photon experiment -- one in a target space of the same radius as classical space, and one in a target space with the reciprocal radius -- they will agree on its duration, and hence on the observed radius of space. Clearly we cannot immediately infer that target space and classical space are one and the same; the remainder of the section explores this situation.

At the start of this section we made our first interpretive decision (to be justified in the next section): we decided that dual theories state the same physical facts. Now that we have clearly distinguished three concepts of space -- classical, target and winding -- we are in a position to describe a \emph{second} interpretive decision, which presents a dichotomy.\\

\noindent \textbf{Interpretation One:} Suppose that the radius of  space has been measured, by the Brandenberger and Vafa experiment say, and found to be very large. Consider a theory, $T$, that sets $R$, the radius of the $x$-dimension, equal to this observed radius. One can then understand $T$ in a naively realist way: take $x$ to represent target space, take the string position/momentum wavefunction $\Psi(x,t)$ to represent a string living in target space, and identify target and classical space. That's a natural way to interpret the theory. But then how is one to understand the dual theory, $T'$, which we are taking to state the very same physical facts as $T$? For instance, $T$ and $T'$ apparently assign different radii to target space, and (for $w\neq n$) apparently assert that the string is wound a different number of times around target space: aren't these physically different states of affairs? A solution is to take the duality mapping as specifying a \emph{translation manual}. From (\ref{eqn:T-D}), in the dual theory, let `$n$' denote the winding number, not wave number, and `$w$' denote the wave number, so that momentum and winding are unchanged! And while in $T$ the $x$-dimension represents target space and the $y$-dimension winding-space, in $T'$ the roles are reversed, so the same wavefunctions pertain to each space as before; and we again identify target and classical spaces so that in $T'$ it is $y$, not $x$, that represents classical space. Finally, as we saw earlier, within each theory the $x$- and $y$-dimensions have reciprocal radii, so in $T'$ the radius of the $y$-dimension is $R/\ell_s^2$: if we understand $T'$ to involve a rescaling of length units by a (dimensionless) factor of $1/\ell_s^2$, then the duals even agree on the radius of target space.\footnote{The factor is dimensionless because the numerator has units of $length^2$. To see that this rescaling is trivial, note that we could have simply have worked in units in which $\ell_s=1$, in which case no rescaling is necessary.} In short, according to this interpretation, duals only appear to be incompatible because they are written in different languages, assigning different meanings to the same words: for example, they appear to assign different radii to target space, but only because they denote different things by `target space'. 

In the framework of first order logic, in this understanding T-duals are related by a permutation of terms that induces a different formal interpretation with respect to a domain with a fixed structure, rather than any change in the domain referred to by those terms. That is, if predicate symbols `$P$' and `$Q$' have extensions $A$ and $B$, respectively, in one interpretation, then in the other they have extensions $B$ and $A$, respectively; and no changes of any other kind.\footnote{For a duality that is not a self-duality, the reinterpretation would replace old terms with new ones, through a `translation manual', rather than permute a fixed set of terms.}

However, such a permutation is trivially possible for any theory with more than one term (of the same kind), so we are left with the question of what distinguishes a (self-)duality from an arbitrary permutation? This question was asked in \cite{Hug:15} (and recently pressed in \cite[\S4]{Wea:19}). The answer suggested there was, roughly, that some theorems, including those whose terms have antecedent empirical significance, are preserved by the duality. For instance, in Brandenberger and Vafa's thought experiment both duals agree on the energy of the particle observed, and the duration of the trajectory: in general, as far as the experiment is described in operational terms, the duals agree on the facts. 

The significance of the invariance of claims stated in terms with antecedent empirical meaning can be brought out by the following example (from \cite{Mot:15}). Consider that if one permutes the meanings of the terms `inside' and `outside', then in the resulting language the Earth is hollow! That is, the solid core is `outside' the surface, and the sky and beyond `inside' -- there may well be aliens living `inside' the Earth! But we antecedently define the `outside' as our direction from the ground, or by the gravitational force, or as the direction of the fixed stars. And the consequence that the moon is outside \emph{in that sense} -- in the same direction as the fixed stars, say -- is not preserved in the new language. On the contrary, the antecedently meaningful, including operational, claims of string theory are common to both duals, so duality is not mere permutation. 

Such invariance is non-trivial but, we now think, does not adequately answer the question of why dualities are physically significant, for permutations of terms are just as trivial, even if some proper part of the vocabulary is held fixed. However, we do not find such triviality as telling against this interpretation of duality. Perhaps the correct understanding is that dualities are physically trivial, and their significance purely formal. So our argument against the adequacy of this interpretation of T-duality will come below, in \S\ref{subsec:fact?}.\\

\noindent \textbf{Interpretation Two:} The second understanding of duality -- which we argue for in \S\ref{subsec:fact?} -- also takes the dual theories as asserting all the same things about the physical world, but now under \emph{a common interpretation} of their terms. In this case  what either says about the physical world must be restricted to their `shared content', in some sense: for instance, the mass spectrum of the string is common to both and hence a physical fact. Similarly, as we saw, the duals predict the same time for a photon to circle the universe: $10^{12}$ years, say. Since the radius of observed, classical space is thus a shared consequence of the duals, it is a determinate, physical fact. 

But the theories do not agree on the radius of target space, nor, as we saw, on what string process corresponds to the photon measurement. Since in this interpretation the terms of the duals denote all the same things, these disagreements are logical incompatibilities between them; and then, because we are taking duals to agree on the physical facts, where the duals disagree, they do not state physical facts. In particular, according to $T$ target space has radius $R$, while according to $T'$, the radius is $\ell_s^2/R$. Thus according to `Interpretation Two' there is no physical fact of the matter which is correct, and with respect to these two values \emph{the radius of target space is indeterminate}. Similarly, it is indeterminate whether the string is wound $w$ or $n$ times around the dimension. And so on.\footnote{\citet[\S6]{Mat:13} argued along similar lines (as we did in \citealp{Huggett:2013sf}), proposing (with misgivings, but without elaboration) that the shared commitments of the duals be thought of as `structure'.  However, his account did not fully recognize the role of derived, classical space in the logic of the situation described in \cite{Hug:15}. The more recent discussion of \cite{MatJoh:18} clarifies this point, and is in agreement with the conclusion of this chapter that (under the assumption of physical equivalence) the spaces of T-dual theories are not strictly identical with classical space, but rather give rise to it.}

The question of the `shared content' of duals is crucial. \cite{Hug:15} suggests that it be understood as any common entailments of both duals. But $T$ entails that target space has radius $R$, and hence that it has a determinate radius; and $T'$ entails that target space has radius $\ell_s^2/R$, and hence \emph{also} entails that it has a determinate radius. So that `target space has a determinate radius' is a common entailment, which contradicts the interpretation of duality that we are currently pursuing! \cite[46f]{Vanoth:80} makes essentially the same point about absolute motion, in an argument against the `syntactic view' of theories, as an adequate approach to the interpretation of physical theories. Drawing the same conclusion here, the natural move is to a more `semantic' view of shared content. Such a position has been worked out in detail in  \cite{Har:19} and \cite{But:19}. Leaving out the details, a theory is understood as a triple of formal states, quantities, and dynamics; such a `bare' abstract structure will generally be realized in a more concrete mathematical framework; then symmetries in general, and duals in particular, are cases of a common bare core theory, with different mathematical representations. We endorse this account, in general terms and specifically, although (for reasons that will become clear later) view it as an idealization in at least some of the cases of interest. (That is not a criticism: we generally view philosophical theories in this way.) We shall return to the bearing of this framework on physical equivalence, but to preview that discussion, the issue for a pair of duals will be whether their differences in representation have physical content, or whether only their shared core does. In the latter, but not former case they will be physically equivalent.

As argued in \cite{Hug:15}, on this interpretation of duality, because the radius of target space is indeterminate while that of classical space is determinate, it follows that \emph{classical space is not identical with target space}. (Similarly, it is not winding space either). Nothing can be both determinate and indeterminate with respect to some property at once. Similarly, it follows that we cannot think naively of strings as spatial objects, since there is no fact of the matter (even in a quantum mechanical sense) of how many times they wrap around a dimension. And so on. As Brandenberger and Vafa conclude (393), `the invariant notions of general relativity \dots\ may not be invariant notions for string theory'.

If this position seems outr\'e, that is only because of the surprising way in which it implements perfectly ordinary considerations. Consider Newtonian mechanics: we know that the predictions of the theory are the same whatever point we choose for the origin, whatever orientation we choose for the axes, and indeed whatever constant state of motion we choose for the frame. And so we think that there is no preferred `centre', that space is isotropic, and that spacetime does not distinguish a preferred state of rest. The fact that our coordinates do distinguish a point, break isotropy, and give a notion of rest is quite clearly an artifact of the representation: inertial coordinates make distinctions beyond those we wish to represent. The same understanding can apply to string theory: T-duality shows that a definite radius for target space and a definite state of winding are not physical, but only artifacts of the representation.

Classical space in this case is therefore derived, or `emergent' from string theory, and in particular from the common core of its dual representations. Let us be very clear that classical space is not `unreal' or `unphysical' for that reason. There are well-known reasons to question the existence of space, but being derived rather than fundamental is not one of them (though the sense in which it approximates a more fundamental physics may bear on the debate).  Space could be perfectly real and perfectly physical, though not fundamental. In general, for all we know there is no ultimate theory of everything, so that everything is derived from something more fundamental. If one insisted that only the fundamental was `real' then for all we know nothing is real -- an absurdity, since we know of many real things! \emph{How} space is derived will be the subject of the next two chapters, but we have already seen how one of its properties -- its observed radius -- can be understood in string theory.\\

So we have two interpretational forks. First, do the two theories describe the same physics, or not? And second, if they do, should we take them literally, with the string living in phenomenal space, and avoid incompatibilities by interpreting their terms differently? Or do they have the same formal interpretation, in which case only their shared consequences are physical? We will work through the first fork with a simple analogy in \S\ref{subsec:D=PE}-\ref{subsec:discusspe}; then turn to the second fork in \S\ref{subsec:fact?}. As we have said, we will argue that the duals are physically equivalent, and that some quantities -- such as the radius of target space -- are not physically determinate.

\subsection{Interpretation: physical equivalence?}\label{subsec:D=PE}


If two theories are dual then \emph{under the duality} the expectation values of all observables are preserved. We emphasize  that `observable' here is used in its physical, quantum mechanical sense, not its philosophical, epistemic sense. That is, observables are the correlates of the system's hermitian operators, generally understood to encompass \emph{all} its dynamical physical quantities, and not merely a proper subset to which we are thought to have privileged experiential access. In other words, saying that `dual observables preserve expectation values' does \emph{not} signify that duals agree only on the values of physical quantities visible to unaided senses, but may differ on those that are not: `observable' means dynamical physical quantity without qualification, and certainly with no implied epistemic privilege.

That understood, systems with dual descriptions need not be physically equivalent: as has often been emphasized in the literature, a formal duality alone is not sufficient for physical equivalence.\footnote{\label{ftnt:SHOrefs}For instance, \cite{Mat:13}, \citet{Hug:15}, and \cite{Har:19} all use the following harmonic oscillator example to make the point.} To explore this issue -- and better understand the equivalence of expectation values -- it is helpful to look at duality in a very familiar system, a simple harmonic oscillator, such as a mass moving horizontally and frictionlessly on a spring, described by the Hamiltonian

\begin{eqnarray}
\label{eqn:SHOH}
H = \frac{p^2}{2m}+\frac{kx^2}{2},
\end{eqnarray}
where $p$ and $x$ are momentum and displacement respectively, and $m$ and $k$ are mass spring constant respectively. This oscillator is dual to another under the duality mapping

\begin{eqnarray}
\label{eqn:SHOD}
\nonumber (m,k) & \leftrightarrow & (1/k, 1/m)\\
(x,p) & \leftrightarrow & (p,-x).\footnote{We will generally say that position is dual to momentum and \textit{vice versa}, although the sign change means that this is not quite accurate. We will pay attention to the sign when it is significant.}
\end{eqnarray}
  (\ref{eqn:SHOD}) is in close analogy to (\ref{eqn:T-D}); position and momentum are the analogues of winding and momentum, and mass and spring constant the analogues of the radius of space.

As for strings, the Hamiltonian and the canonical commutation relations are the same under the duality (for the latter, $[x,p]=[p,-x]$). By the same logic then, the expectation values for all pairs of dual observables agree, so that if any series of values of the quantities represented by dual observables is consistent with either oscillator, then it is compatible with both. However, the dual theories can clearly be used to describe two distinct concrete, physical oscillators in our world, one with mass $m$, and one with mass $1/k$ (unless $m=1/k$). No one questions that these would be dual, \emph{but physically distinct oscillators.} Nevertheless, in this subsection we argue (with many others)  that one should draw the opposite conclusion in the parallel case of T-duals differing in the radius of space. We will explain why by further unpacking our example.

In particular, we need to consider carefully the \emph{measurements} that might distinguish the two oscillators. As with any symmetry, we are interested in the question of whether indistinguishable systems are physically identical, so  we have to understand clearly what can and cannot be distinguished. In the first place, given concrete oscillators we could simply dismantle them and place their bobs on a scale to determine the masses, and thereby distinguish them. But this is not helpful to our enquiry into T-duality, because there is no analogous experiment that could determine the radius of space, only Brandenberger and Vafa's equivocal experiment, and its ilk. If we thus don't have direct empirical access to constant, classical, c-number parameters like mass, radius (or spring constant), then the question is whether duals can be distinguished by measurements of their dynamical, quantum observables. (Recalling our discussion of `observables', we bear in mind that we are assuming the measurability of all such quantities, not a just proper subset accessible to human senses.)

For instance, the quantum harmonic oscillator energy spectrum is $E_n = \hbar\sqrt{\frac{k}{m}}(n+\frac{1}{2})$, so observations of the energy can determine the c-number $k/m$. But such measurements clearly cannot determine whether $\langle \mathrm{mass},\mathrm{\ spring\ constant}\rangle = \langle m,k \rangle \mathrm{\ or\ } \langle 1/k,1/m \rangle $, since they agree on the ratio of mass to spring constant. In general, measurements of observables that are invariant under a duality will (obviously) not distinguish the duals. And some observables will be invariant: at least the energy, since the Hamiltonian must be invariant to preserve the duality over time.

But not all. According to (\ref{eqn:SHOD}) $x$ in one dual agrees with $p$ in the other, not (in arbitrary states) $x$ in the other; and similarly for $p$ and $-x$. The duality (like T-duality) does \emph{not} assert that both theories assign the same values to the same mathematical objects (operators and their expectation values), but rather that they instantiate the same `pattern' of values. For instance, imagine a table of pairs of measured values at a series of times; if they agree with one oscillator's $x$ and $p$ expectation values at those times, then they equally agree with the expectation values of $p$ and $-x$, respectively, for the dual oscillator: at any time, the position of one is numerically equal to the momentum of the other, and the momentum of the first to minus the position of the other. The situation is exactly the same as for a string, in which the values of winding and momentum are exchanged by T-duality, as we explained and as figure \ref{fig:tdual} illustrates. And similarly for other dualities.


We agree with \cite{Wea:19} that dualities are thus formally distinct from instances of empirical equivalence as usually characterized.\footnote{We don't agree with his suggestion that the duality literature misses this point (see footnote \ref{ftnt:SHOrefs}), though his critique has prompted us to explain it more carefully. He uses the simple example of source-free electromagnetic duality in much the way we and others have used the oscillator.} As in (\ref{eqn:T-D}) and (\ref{eqn:SHOD}), a duality is an invariance under a \emph{mapping} between observables; a permutation of observables if the theories are self-dual (as in T-duality), or a correspondence between distinct sets of observables if not (as in the `holographic' duality discussed in \S\ref{sec:GGD}). It is not simply an invariance of some quantities when others are transformed or `translated' to new values (though some observables will be preserved by a duality). However, Weatherall's approach does not adequately recognize how the issue of empirical equivalence turns on the question of which quantities have independent physical significance, and which obtain their significance through the dual theories themselves. 

So suppose again that one is given the table of oscillator observation pairs: and suppose they agree with the expectation values of $x$ and $p$ for some oscillator, hence for $p$ and $-x$ for its dual. Applying the standard convention that $x$ represents position and $p$ momentum to \emph{both} duals (so adopting our Interpretation Two), knowledge of which column describes position measurements and which momentum allows one to distinguish the dual oscillators. Once again, if $\langle$position, momentum$\rangle=\langle i,j\rangle$ for one dual, then $\langle$position, momentum$\rangle=\langle -j,i\rangle$ for the other, and these are generally unequal, since the oscillators move differently. This seems to be the normal case, in which of course the dual oscillators are empirically distinguishable by position and momentum measurements.

But the same pair of columns with \emph{no} indication of which is position and which momentum, do \emph{not} distinguish the duals: $\langle i,j\rangle$ might represent $\langle$position, momentum$\rangle$ or $\langle$momentum, $-$position$\rangle$. In the actual world, of course there would still be a fact about which column was \emph{really} the result of position measurements, and which momentum. The situation could only occur if, say, a careless lab assistant neglected to label the columns when recording the data; though then the inability to discern duals would only be epistemic. But what if the very meanings of `position' and `momentum' were called into question, raising the question of what exactly the two columns refer to?

This cannot happen in the normal case, because position and momentum are well-defined by a theoretical and experimental framework independent of the harmonic oscillator: especially, their general theoretical understanding in classical and quantum mechanics, the models of very many other physical systems in which they are dynamical quantities, and the numerous techniques for their measurement. For instance, dual oscillators can be distinguished by weighing their masses. Or by coupling them to some external system that does not respect the duality, but depends directly on the position: reflecting light off the bob, say. Or by `looking inside' the oscillator in some other way.

Calling into question the distinction between the dual quantities means ignoring all of that, something that can only be done by supposing a world in which none of that framework surrounds the oscillator, so a situation in which `position' and `momentum' are not `externally' meaningful: in other words, a world in which there is nothing but a single oscillator. Then there is no weighing or shining a light on the bob, since there are no scales or light; the difference between oscillator position and momentum no longer makes a difference to other systems, because there are none. No physical operations `look inside' the oscillator.

Equation (\ref{eqn:SHOH}) can be taken to describe such a world. One then  reflexively imports the usual interpretations of `$x$' and `$p$' as position and momentum in the senses given by the actual world framework. But then one discovers (\ref{eqn:SHOD}), and that the pattern of values for these quantities would be the same under the opposite interpretations of `$x$' and `$p$'. And so the question is which of these two identifications of the quantities $x$ and $p$ is the correct one in the single oscillator world in which  (\ref{eqn:SHOH}) is the \emph{complete} physical theory? Is the property of the lone oscillator denoted `$x$' position or momentum in our familiar sense? We argue that under the given circumstances there is no fact of the matter: nothing internal to the oscillator world need determine how it instantiates the qualities of our world.\footnote{If you like, we advocate a kind of second-order antihaecceitism for qualities: dual worlds can't differ simply in how dual qualities are instantiated. We believe that such a view is compatible with Lewis' nominalism, including some version of his natural properties view, for instance.} (Clearly our free choice of the symbol `$x$' to label position in our world carries no such ontological weight.) But that is to say, in theory terms, that the duals describe the same world; that they are physically equivalent, that there is just no `inside' to see.

Further, we argue, the case is just as in string theory, taken as a theory of \emph{everything}. Consider a world in which there is only a string in a spacetime with a closed dimension: now $p$ (or $n$) and $w$ are the duals (\ref{eqn:T-D}), in analogy to $x$ and $p$ for the oscillator. We know that dual systems agree on the energy, and through the analysis of Brandenberger and Vafa, radius measurements. Such measurements would in principle allow one to determine the values of \emph{all} observables, including momentum and winding. The issue is not the unobservability of momentum or winding, but rather that of determining \emph{which measured quantity is position, and which is winding}. That is, the duality preserves the pattern of observables: some are invariant, and others permuted -- if $\langle n,w\rangle = \langle i,j\rangle$ in one dual then $\langle n,w\rangle = \langle j,i\rangle$ in the other. The pattern itself will not distinguish the duals, and there is no broader theoretical framework in the lone string world to settle which value is $n$ and which $w$. By stipulation, there is no broader external theory in which strings with T-dual assignments of $n$ and $w$ are not dual, allowing the duals to be distinguished. As with the oscillator, there is no physical way to `look inside' the system to see which dual it is. It seems that we should draw the parallel conclusion: no preferred identification of the quantities with those of the actual world, and no physical difference between the duals. That the duals describe the same world, and that there is no `inside' to see.

But what of the `low energy' limit of the theory in which something like quantum particle physics is found? In that framework particle momentum is well-defined, and observable. But the lesson taught by Brandenberger and Vafa is that the state of a particle, or the measurement of that state always have dual analysis: the state of a stringy particle can be understood equally well in terms of the target space state of a string, or the dual winding space state; and any measurement of particle momentum in terms of dual target and winding space processes. The measurement of the radius of space is just one example of something general: low energy physics cannot break a duality, since any `reduction' has a dual, hooking up a single low energy structure to either of two dual high energy structures.

We also want to head off the line of thought that we can just see -- immediately experience -- that the radius of space is large, and that things would seem different if it were not. Brandenberger and Vafa's argument applies here. Given that our visual experiences supervene on the physical, whatever physical process that underwrites our experience of a large dimension is realized in both duals: in one as a process involving momentum modes, say, and in the other involving winding modes. We have been arguing that we should take these to be different representations of just one process, but even on the view that counts them as distinct physical possibilities, a fairly mild assumption will guarantee the indistinguishability of the duals even in direct experience. For the two processes will only be experientially distinct if visual experiences depend on the processes grounding them involving spatial (not winding modes): that T-dual brains are not identical minds. It is, in other words to privilege the spatial in the physical theory of mind. But we see no particular motivation for such a view: rejecting it means that dual brains have the same experiences, so that things would not appear any different at all if target space had the reciprocal radius. Hence we cannot just `see' which of the two possibilities holds, and considerations of direct experience provide no reason to think that there are two physical possibilities at all.

We therefore conclude, in parallel with the lone oscillator, that in the lone string world there is no fact of which of $p$ and $w$ is momentum and which is winding, and that the duals are thus physically equivalent. Moreover, if string theory is understood as a theory of everything, then whether there is one string or many makes no difference to the argument, and so we also conclude that T-duals of full string theory are physically equivalent. Our conclusion is not a logical necessity, nor do we think there are compulsory  semantic or ontological principles that can force the conclusion that dual theories of everything describe the same physical possibility. But the case of dual total theories is clearly one in which the putative differences are `hidden' in a very strong sense -- a unique mass is impossible to determine from the physical quantum quantities of the harmonic oscillator, just as a rest frame is from relativistic quantities in special relativity. And when there are quantities that do not supervene on any of the other physical quantities, and when there is no reason to think that different values for them can be determined directly, then at least from a practical, scientific point of view, it makes sense to treat those differences as non-physical (until some new, well-supported theory shows how they are connected to physical quantities). In other words, long established, well-motivated scientific reasoning should lead us to think that dual total theories represent the same physical situation. 

Of course one now wonders what physical equivalence under duality really amounts to. What exactly is being claimed? Well, we have already described two interpretations of the claim in the previous section. In  \S\ref{subsec:fact?} we will turn to the question of which we favor and why. Before that we will discuss the general claim of equivalence of duals, and recent philosophical reflection on the conclusion.

\subsection{Interpretation: the meaning of physical equivalence}\label{subsec:discusspe}

Our line of thought was presented in \cite{Hug:15} (\cite{DieDon:15} argue similarly), but since then other authors have clarified or questioned the conclusion of physical equivalence.\footnote{Philosophers have focussed on the issue of physical equivalence, and the implications of the common core for emergence, but \cite{Daw:17} emphasizes that string theorists have found duals a useful tool to gain different perspectives on the underlying string theory. Indeed, he argues that this is their main significance; we agree with the importance of the use he describes, but claim that quotienting is the important implication for understanding the topic of this book -- spacetime emergence.} First, \cite{Har:19} (building on a series of earlier papers cited there, including that with Dieks et al. just cited, and \cite{De-But:19}) develops this idea more formally and thoroughly. Summarizing, in de Haro's terms, dual oscillators are distinct because in our world, their common core can (in a precise sense) be `extended' to -- embedded in -- a larger theory that gives `external' meaning to their terms: mass, spring constant, momentum, position. But in a world in which the common core instead describes \emph{everything}, then the duals are nothing but different tools for computing the dynamics, with their differences (in $m$ and $k$, and $x$ and $p$) as nothing but empty conventions used to turn the mathematical handle. In that case there would be no larger theory of the world (without uninterpreted surplus structure) in which the core could be embedded; it is `unextendable' to a more comprehensive theory, and hence cannot receive an external interpretation. Instead it can only have an `internal' interpretation: possible states are fully distinguished by the value-pairs, have the same allowed histories, and are interpreted as the values of the only two physical quantities of the world. As we noted above, we endorse this picture, but we think it idealizes the situation: as we will discuss shortly, we need not have an explicit formulation of the common core in order to know something of the shared physical content of duals.

Given de Haro's framework, the question of physical equivalence has two parts. First, could we ever reasonably believe that the common core of a pair of duals was not extendable? That it captured all the physical structure of the world in its domain, so that it was not just part of a broader (perhaps more fundamental) theory? Generally, unextendibility will not be a purely formal property of a theory, but will depend on its intended application: real world oscillators are extendable, but one can arguably stipulate a world in which they are not. So a typical way to frame the question will be regarding its intended application to the actual world. Then one may want to apply methodological principles such as ontological simplicity to move from duality to unextendability, and thence to an internal interpretation. For instance, if the world constantly manifests Lorentz symmetry, why postulate some unknown physics that picks out a rest frame? Alternatively, de Haro suggests that physical principles of a theory will be used to answer the question: perhaps in this case the Lorentz symmetry of the theories. But in a sense this approach will also rest on methodological principles, for how else are we to decide the physical principles themselves?

Second, suppose that the world were such that the common core of a pair of duals indeed has no external interpretation: does it follow that the duals are physically equivalent? Perhaps instead they could describe a pair of worlds in which different physical quantities are instantiated in isomorphic patterns. This question will in part depend on considerations from philosophy of language: does an interpretation of the core provide a relation between a single world and both duals, so they refer to a single domain? De Haro shows that fairly mild assumptions about reference justify such a conclusion. However, one might ask whether it is possible for there to be two interpretations of the core, each relating the duals to different worlds. In such a case though, any one interpretation of the common core will serve as an interpretation of the duals, since only the core has physical significance. And so, de Haro points out, any one interpretation will relate all duals to the same world. Hence, even if one interpretation is `better suited' to the mathematical representation of one dual than another, this will only be a pragmatic matter, not a semantic one that could produce inequivalence.

Positive answers to both questions (Unextendable? Unequivocal internal interpretation?) for a pair of duals means that they are physically equivalent, with their content exhausted by a single internal interpretation of their core. It should be clear that such a conclusion, therefore, does not follow simply from some formal property of theories, but is a matter of interpretation and philosophical theory, something emphasized by \cite{But:19}. However, some recent commentators have raised substantive issues regarding physical equivalence, to which we would like to respond. 

Both \cite{ReaMol:18}, and \cite{But:19} discuss significant cases resembling T-duality, in which a profound symmetry relates two theories, arguing that an inference of physical equivalence is \emph{not} thereby justified. 
For on more careful consideration, there is an important difference between Lorentz invariance and T-duality, namely the \emph{knowledge} of an underlying formal structure which unifies different frames: a common core that makes explicit the unphysical surplus structure introduced when a frame is chosen. (Unphysical in the sense that a convention is involved, even though that convention will have to refer to physical objects to pick out an origin, orientation, and so on.) In the case of T-duality there are just the duals, and no known explicit common core; that would be `M-theory', an exact completion of string theory (something that will be a recurring topic in the remainder of this part of the book). Does this make a difference to claims of equivalence? Read and M\o ller-Nielsen, and Butterfield think so. Consider the related example of (full) Newtonian spacetime versus Galilean (or `neo-Newtonian') spacetime (e.g., \cite[chapter 2]{ear:89b}). Suppose one knew Galilean symmetric physics but only of Newtonian spacetime. Would one be justified -- just from Galilean symmetry -- in inferring the physical equivalence of two `theories' that differed only in the standard of absolute rest? Or should one believe these to be inequivalent states of affairs? (Or be agnostic?) \emph{Once} one discovers Galilean spacetime it is reasonable to take that to properly capture the geometry of spacetime, but until then? After all, that is the analogue of the situation with string dualities.

It seems that everyone is agreed that things are not clear cut. But Butterfield says that inequivalence is a reasonable option (\S1.2); and Read and M\o ller-Nielsen say that inferring equivalence is not justified (\S3.3). While we think that although it is most reasonable to infer equivalence for T-duals, we acknowledge that the point is open to debate, and others could judge differently. We do not see the availability of an explicit formulation of the common core as particularly important for inferring equivalence; we believe it to be cogent to assert that the physical content of duals is `that which they say in common', and argue that making explicit  this common content is not necessary in order to accept the assertion.\footnote{There is a recent literature debating whether Ramsifying -- here that `there are $x$s such that they satisfy the common commitments of the duals' -- is reasonable. For instance, \cite{Dor:10} argues against this move in a number of instances, while \cite[chapter 5]{Sid:} defends it (in some cases). We find that there are rather strong intuitions on both sides; it is a topic we are happy to see being explored more carefully.} 

In particular, our intuitions about the right thing to say about the Newtonian spacetime example run in the opposite direction to Butterfield, M\o ller-Nielsen, and Read. They think that one would best assume a standard of rest until Galilean spacetime is discovered; we think that one should conclude that  apparently different ascriptions of rest are in fact equivalent. (Which is not to say that we fault Newton in historical context for accepting both absolute space and Galilean relativity; we make our judgement from a historical perspective that has accumulated a great deal of additional understanding of logic, semantics, mathematics, nature, and science.) And while agnosticism about equivalence is the more epistemically \emph{cautious} course, we don't believe it to be more epistemically \emph{virtuous} for that. What we think is that global theories -- such as Newtonian gravity -- do have a good track record for turning out to be unextendable, and that in particular string theory is promising as a complete unified theory in its domain, and so is reasonably thought to be unextendable. And from that we do think physical equivalence is the reasonable conclusion.

Of course, we agree that giving an explicit formulation of a common core is a significant goal even when one has accepted physical equivalence; it would be an explicit formulation of the physical content of the duals, and so crucial to fully understanding them. Indeed, the expected utility of finding such an explicit core will (all things being equal) be greater the more likely one thinks duals are equivalent; the more likely that is, the more likely it is that an explicit core exists. So we don't find attractive Read and M\o ller-Nielsen's (\S3.3) `motivationalist' position: that one is most rational only to accept equivalence once an interpreted common core is known, and that one should seek it. As we just explained, we find sufficient reasons to accept the equivalence of T-duals, and that acceptance is (additional) motivation to seek the common core. Indeed, our discussion of Brandenberger and Vafa, and our investigations in later chapters contribute to such a project. As does (with impressive results) the work of de Haro and his collaborators (in addition to work already cited, see \cite{DieDon:15,De-May:16,HarTeh:16,Har:17}). 

All that said, it may be surprising that we have some skepticism that the common core of string duals can be formulated in a closed, complete formalism. The reason is that string theory as currently formulated is an essentially perturbative theory: as we shall see in detail in the following chapters, one postulates a classical limit of some as yet unknown theory, including a classical spacetime, and then studies quantum perturbations around it. As such, it is not clear that there will be some way of completely describing a shared core structure better than `that which the duals have in common'; the duals themselves are inherently limited as descriptions of the world. In this case, de Haro's framework is just an idealization, as we suggested above. This situation is compatible with finding out specific aspects of the common core, as we have just described, but incompatible with stating a closed, complete interpretation. Instead, one hopes that the content of the string duals will be found within a theory to which they are the perturbative approximations, namely `M-theory'. (In addition to Read and M\o ller-Nielsen, and Butterfield, this point is stressed as a motivation for studying duality by \cite{Le-Rea:18}.) More than likely though, M-theory contains both more and less content than either of the duals, and so is not the same as simply quotienting them. 

One might then ask what the value is of studying the duals, but this is a question we have addressed a number of times: we are seeking to understand the fragments of existing quantum theories for ways in which existing concepts of space and time might be modified in a successful theory. So our attitude towards the investigation of duals has the same spirit. Moreover, the need to interpret a perturbative quantum theory is nothing new; just the same policy has been fruitfully pursued for quantum field theory (something defended in \cite{Wallace_2006,WALLACE2011116}). However, we should acknowledge that in chapter \ref{ch:N-4} the pertubative nature of the duals will limit our investigation.

\subsection{Interpretation: factual or indeterminate geometry?}\label{subsec:fact?}

We will proceed on the understanding that T-dual theories describe the same physical situation. The question now is \emph{what} situation that is, in particular with respect to the geometry of space. Above we described two possibilities: it could be that the duals agree that the radius of target space is greater than $\ell_s$, and the apparent inconsistency is resolved by understanding duality as a permutation of terms, a relabeling. Or it could be that the duals should receive the same formal interpretation, so that only their common pronouncements describe what is physical: for instance, a unique radius to phenomenal, but not target, space. In this section we will make a couple of brief comments on the two possibilities, and then explain why we favor the second.

Talking of `relabeling' the terms of a theory may suggest that the difference is between passive and active interpretations of duality. But that clearly isn't correct: an active transformation links two distinct states of affairs, but both interpretations agree that there is only one possibility, so neither amounts to the view that T-duality is an active transformation. Moreover, T-duality cannot be seen as a passive transformation in the sense that the duals are descriptions of a single situation from two points of view, for the duality does not map `observers' or concrete `reference frames' into distinct but symmetrical observers and frames. And in the looser sense that both interpretations take duals to be distinct representations of the same physical situation, both interpretations take a duality to be equally `passive'.

In fact, the two interpretations that we have described are much closer to the interpretive options that arise in the case of a gauge symmetry. On the one hand, maybe there is `one true gauge' \citep{Hea:01}: in the present context, phenomenal space is identified with target space, and has a definite radius $R>\ell_s$. On the other, maybe apparent differences in choice of gauge are nothing but differences in `surplus representational structure' \citep{Red:75}: target space is distinguished from phenomenal space, and the difference between target spaces of radii $R$ and $\ell_s^2/R$ is merely a difference in representational fluff. We won't pursue this parallel to gauge symmetry in field theory at length, but a couple of points are worth making. First, duality is neither a local nor a continuous symmetry of the kind found in field theory, so much of the philosophical discussion of those theories is inapplicable. Second, that said, at $R=\ell_s$ there is a continuous SU(2)$\times$SU(2) gauge symmetry of which T-duality is a part (e.g., \citealp[247-8]{Pol:03}). Thus, in this sense at least, T-duality is formally, and not just conceptually, a gauge symmetry.\footnote{See \citet{Hea:07} and the responses to it for continuous gauge symmetries in general. The SU(2)$\times$SU(2) symmetry entails that an infinitesimal increase of the radius from $R=\ell_s$ is the same as an infinitesimal decrease. \cite{Read:2014fk} makes a related comparison, but between string dualities and diffeomorphism symmetry rather than conventional gauge symmetries.}

So, why do we advocate the indeterminate $R$ interpretation? After all, the definite radius view presented above is intuitive, in that it says that strings live in a space with an observed radius $R>\ell_s$; whether that space is called target or winding space. However, there is a \emph{distinct}, indistinguishable definite radius view  according to which strings live in a space whose radius is $\ell_s^2/R$; whether that space is labeled target or winding space! Generally, if there is one true gauge, then there are as many distinct possibilities for it as choices of gauge: in this case two, depending on the radius of the space in which the strings literally live, move and wind. According to one choice, the space of experience is the one in which strings live, while according to the other the space of experience is much bigger than the one in which they live: from Brandenberger  and Vafa we understand that the same appearances arise from a string's momentum state in one dual, and from its winding state in the other. The bottom line is that understanding T-duality as a mere permutation of terms leaves open what underlying facts are equally described by the duals, because such an understanding is compatible with different true gauges. Hence `Interpretation One' does not really address the issue it was supposed to resolve: dual theories are physically equivalent on this interpretation, but there is a second pair of duals that differs physically from the first, but \emph{only} with respect to an unobservable radius. If one is satisfied with that situation, then why was one not satisfied with physically inequivalent duals?

Moreover, these considerations point to an analogy to related cases in which we usually do accept that there is no fact of some matter (we alluded to a similar example earlier). For instance, one could claim that there is a preferred rest frame in spacetime, even though it has no physical significance in special relativity.  One could even claim that it is some frame which can be picked out physically and phenomenally: for example, perhaps the fixed stars (idealized as an inertial frame) are at rest. This proposals will strike most readers as completely unmotivated. But replace `frame' with `radius', and the fixed stars with the observed radius, and the parallel is perfect. Looked at this way, the definite radius view appears as a reactionary attempt to preserve aspects of an old theory when it is superseded, and understood as merely effective.

However, since \cite{Hug:15} we have recognized a way of defending a version of Interpretation One that avoids these objections.\footnote{We are especially grateful to Neil Dewar for a useful discussion of the following.} Formally, the common formal core of T-duals (in our toy model) is a pair of spaces, one big and one small. There is nothing indeterminate about the radii of the bigger and of the smaller ($R$ and $1/R$, respectively, in $\ell_s=1$ units); and low energy phenomena are understood in terms of states in the bigger, because it allows longer wavelength, lower energy, wavefunctions. (In figure \ref{fig:tdual}, delete `space' and `winding space', and the only difference between the upper and lower figures is the trivial coordinate relabelling $x\leftrightarrow y$.) One could interpret the duality as showing that there is no more physical content to the theory than this. Especially, as showing that the \emph{distinctions between} the terms `target space' and `winding space', and \emph{between} the related `momentum' and `winding' of the string, are \emph{without physical content}. For as soon as these have independent meaning, we can distinguish the two duals; even if we recognize them as different descriptions of the same state of affairs. In short, this interpretation means that talking of a `string' in any classical spatial sense at all evaporates, because all one has is some quantum object, described by a product of wavefunctions, one in each space. 

One is of course then free to adopt the \emph{convention} that `target space' simply means `big space', and `winding space' simply means `small space', and that `momentum' and `winding' refer to wavefunctions in big and small space, respectively. That is to strip the terms of any of the physical content with which they were introduced at the start of this chapter. So equally, one could have adopted the opposite convention, and declared `winding space' synonymous with `big space'. Either way, these are just different terms one might choose for the same concept. Then one can preserve Interpretation One, understanding the permutation of terms not as a choice of `true gauge', but as a mere linguistic convention. (Similarly, saying whether space is target space or not simply reports the convention.) But adopting either convention adds nothing to the theory, since `big' and `small' would do just as well; their only significance is heuristic, as reminders of a certain way to derive predictions.

In effect, this `two space' interpretation extends the familiar idea (investigated in \cite[chapter 3]{ear:89b}) that the geometry of a spacetime should have the same symmetries as the physical laws: in particular, if a spacetime has additional symmetry-breaking structure, it is preferable to find a spacetime which `quotients' it away. The proposed new formulation of string theory quotients away the distinction between target and winding space, leaving only `big' versus `small' as physically meaningful, so that the symmetries of the theoretical representation match those of the dynamical physical quantities. It agrees that space is not target space, if `target space' is supposed to denote something more than `big space'. But it does not admit three distinct spaces, since space is not emergent, but identified with the larger of the two string spaces: the one to which we refer spatial phenomena.

While we endorse Earman's prescription in general, and find this interpretation appealing, we do not think that it is the correct account of string dualities. The simple quotienting in this case is an artifact of the particular formulation and duality. Our `double field' approximation leads to two spaces with interchangeable roles, whose initial difference can then be ignored. But in other dualities, introduced in \S\ref{sec:beyondT}, things are not so simple, and a quotiented formulation in which space is clearly identifiable is not available: dimensionality or other global topology changes, or weak couplings  are exchanged with strong ones, or even the elementary with the composite. But without such a common core, with a structure identifiable as space, the approach breaks down.\footnote{We want to thank Keizo Matsubara for emphasizing especially this point to NH in their many discussions about duality.} Then, on the one hand, parity of cases implies that T-duality is correctly understood in the same way; on the other, insofar as we are taking our toy model of T-duality as a model of all dualities, we need to interpret it the same way (regardless of whether that is correct for the case in hand). 

Even in the case of T-duality in a realistic perturbative string theory (as opposed to our toy system), formally a string will live in three large spatial dimensions plus six or 22 microscopic dimensions (plus time); and from the worldsheet perspective, momentum and winding will correspond to distinct sets of quantum field excitations. The microscopic dimensions are beyond normal observation, but are still `big' on the string scale ($r\ll\ell_S$), so larger than winding space; however small enough that that at energies low on the string scale, both momentum and winding will contribute to string processes. The model we have worked with collapses all this structure into a pair of wavefunctions, but once it is put back, it is not clear even in this perturbative case how the quotienting strategy can be applied. We will return to this question briefly in \S\ref{sec:openT}, when we will see that open string duality makes this simple quotienting even less satisfactory.

In other words, while we agree that the proposed two space interpretation is reasonable for a system in which our double field model is the exact, complete description (as we treated it in comparison with the harmonic oscillator), when it comes to interpreting string theory we have to bear in mind that the model is an approximation. The proposed quotient depends on that approximation and, we argue, does not apply to string theory more fully; hence the two space interpretation is not a viable account of string theory. For string theory then, we endorse Interpretation Two instead. That is, target space (like winding space) has a radius indeterminate between $R$ and $1/R$, and so cannot be identified with classical space, which has a determinate radius of $R$ -- in other words, space is emergent.\\

\section{Beyond closed string T-duality}\label{sec:beyondT}

This chapter focusses on the technically simplest example of string duality, but we should briefly survey the other important cases. The philosophical issues that arise are -- we claim -- the same as for closed string T-duality, though because questions of equivalence are not separable from questions of interpretation, the conclusions need not be the same (though we are inclined to say they are). Here, however, we just focus on sketching the relevant formal ideas, and leave aside such questions. 

Below we will discuss T-duality for open strings (\S\ref{sec:openT}), which must be treated differently to closed strings, since they cannot enclose a dimension; and the `holographic' duality between strings in a gravitational field and a gauge field on the boundary of  spacetime (\S\ref{sec:GGD}). Brief as those treatments are, two other dualities will only be mentioned in passing (introductions and investigations can be found in \cite{Ric:11}). 

First there is `mirror symmetry', which states the equivalence target spaces of different topologies. More specifically, this equivalence holds for supersymmetric strings, which we saw live in a 10-dimensional target space. If it takes the form of 4-dimensional Minkowski space, plus 6-dimensional closed `compact' dimensions (as one might model our universe), then mirror symmetry holds between specific pairs of compact dimensions of different topologies. Insofar as these are taken to literally represent space then according to our preceding analysis there is no fact of the matter of the `shape' of space. (Though one could argue instead, with \cite{MatJoh:18}, that that such indeterminacy sometimes indicates that the compact dimensions are not spatial at all, but rather represent internal degrees of freedom.) While our conclusion that space has no definite classical topology is even more conceptually dramatic than our earlier one that spaces of definite topology have no determinate radius, mirror symmetry is in fact a mathematical generalization of T-duality (\cite{StrYau:96}). Mirror symmetry is especially interesting because its discovery sparked much of the initial excitement about dualities when it was used to simplify and solve some difficult mathematical problems.

Second, there is `S-duality', which also has applications in mathematics. This relates different `types' of superstring theory, characterized by their different boundary conditions. In this regard S-duality is analogous to open string T-duality, to which we will shortly turn. However, it is also characterized by exchanging strong and weak values for the string coupling strength: for instance,  Type I string theory with a strong coupling is dual to SO(32) type with weak coupling (while Type IIB has a self-duality between strong and weak couplings). S-duality has been less discussed by philosophers, but is thought significant for providing an important clue that the different supersymmetric string types are different perturbations of a single underlying, M-theory.\footnote{Other interesting dualities, and some profound insights into the nature of duality in general can be found in \cite{Pol:17}. Especially, he discusses elementary-composite duality, also investigated by \cite{Cas:17}.}

\subsection{T-Duality for open strings}\label{sec:openT}

On the face of things, it is hard to see how T-duality could be extended to open strings. The winding number of a closed string is conserved classically or in the absence of string interactions because the topology of target space prevents the string from being contracted to a smaller winding number. But an open string can (topologically) always be contracted to a point, in any space, so does not seem to have a winding number. But in that case, one cannot exchange momentum and winding, and (\ref{eqn:T-D}) does not seem to apply at all. (Or if you prefer, $w$ is always zero, so T-duality must fail for non-zero momentum.) However, T-duality does apply to open strings, in a surprising and very important way, which illuminates the physical content of the theory. We cannot see this at the level of the toy double field approximation we have used so far, but need to draw on the theory developed in chapter \ref{ch:N-1}.

First we will work out how to implement T-duality for closed strings in those terms, and then we   apply the transformations to the open string.\footnote{\label{ftnt:ostdrefs}The following treatment follows \citet[\S8.3, 6]{Pol:03}, \cite[\S6.1]{BecBec:06}, and \cite[chapters 17-18]{Zwi:04}.} We start again with the equation for a string wound around a closed dimension (\ref{eq:wound}), ignoring the vibrational part for simplicity for now; what follows can readily be seen to hold for that part too. 

\begin{equation}
\label{eq:N2closedstring}
X(\tau,\sigma) = 2\ell_s^2p\tau + 2Rw\sigma = 2\ell_s^2n\tau/R + 2Rw\sigma, 
\end{equation}
using $p=k=n/R$ (since $\hbar=1$). Next we expand this into left and right moving pieces:

\begin{equation}
X = (\tau+\sigma)\cdot(\frac{\ell_s^2n}{R} + Rw) + (\tau-\sigma)\cdot(\frac{\ell_s^2n}{R} - Rw) \equiv X_L(\tau,\sigma) + X_R(\tau,\sigma).
\end{equation}

We observe now that if we take the rescaled reciprocal radius, $R\to\ell_S^2/R$, then:

\begin{eqnarray}
\nonumber X_L & \to & (\tau+\sigma)\cdot(\frac{\ell_s^2w}{R}+Rn)\\
X_R & \to & -(\tau-\sigma)\cdot(\frac{\ell_s^2w}{R}-Rn),
\end{eqnarray}
so that if one also interchanges $w\leftrightarrow n$, then $X_L\to X_L$ and $X_R\to -X_R$. Then summing the transformed left and right moving parts, we find:

\begin{equation}
\label{eq:N2Tdual}
X\to 2Rw\tau  + 2\ell_S^2n\sigma/R.
\end{equation}

Comparing this expression with (\ref{eq:N2closedstring}), paying attention to the timelike and spacelike coordinates, we see that it describes a closed string with wavenumber $w$ and winding number $n$, but in a target space of radius $\ell_S^2/R$. That is of course what we expected from our previous quantum analysis, (\ref{eqn:T-D}). In other words, in this formalism we implement closed string T-duality with the transformation:

\begin{equation}
\label{eq:ostd}
  X_L\to X_L \qquad\text{and}\qquad X_R\to -X_R.
  \end{equation}
We postulate that the transformation remains unchanged for the open string.

So we start with the general equation of motion for the open string (\ref{eq:open}) (recalling footnote \ref{ftnt:rescale}):

\begin{equation}
X = 2\ell_s^2n\tau/R + i\sqrt2\ell_s\sum_{j\neq0}\frac{1}{j}\alpha_j e^{-ij\tau}\cos{2j\sigma}
\end{equation}
For the open string it is illuminating to leave in the part of the solution describing vibrations, as we shall see. Break this into left and right moving parts, $X=X_L+X_R$,

\begin{eqnarray}
\nonumber X_L & = & \frac{\ell_s^2n(\tau+\sigma)}{R} + \frac{i\ell_s}{\sqrt2}\sum_{j\neq0}\frac{1}{j}\alpha_j e^{-ij\tau}e^{-i2j\sigma}\\
X_R & = & \frac{\ell_s^2n(\tau-\sigma)}{R} + \frac{i\ell_s}{\sqrt2}\sum_{j\neq0}\frac{1}{j}\alpha_j e^{-ij\tau}e^{+i2j\sigma},
\end{eqnarray}
and apply our T-duality transformation (\ref{eq:ostd}) to obtain the dual state:

\begin{equation}
X_L-X_R = 2\ell_s^2n\sigma/R + \sqrt2\ell_s\sum_{j\neq0}\frac{1}{j}\alpha_j e^{-ij\tau}\sin{2j\sigma}.
\end{equation}
What does this T-dual state represent? 

First, we notice that there is no term linear in the time, so no overall linear motion; all the string's motion is vibrational. But if there is no linear momentum term, what does $n$ now represent? As in (\ref{eq:N2Tdual}), the term linear in $\sigma$ has the correct form for a string wound $n$ times around a space of radius $\ell_S^2/R$. But how can an open string have a meaningful winding number? To answer this question, consider the ends of the string, $\sigma=0,\ \pi$. Because of the $\sin{2j\sigma}$ term, they have no time dependence at all -- they are fixed in space! (Or rather, they are fixed in the dimension in question.) That is, they satisfy the Dirichlet boundary conditions that we noticed in (\ref{eq:sigmin}). So that is why an open string can have a sensible winding number: if its end are attached to something, it can no longer be topologically shrunk to a point! So the physical interpretation is that the state T-dual to a freely moving open string in a space radius $R$, is an open string wound around a space of reciprocal radius, but with its ends fixed in place. (And it's easy to see that the ends will be a distance $2\ell_S^2\pi/R$ apart.) Of course, one has to check that the transformation really is a duality, that these two very different strings do agree on expectation (values under the duality), as for closed string. But they do (as you can see in the references in footnote \ref{ftnt:ostdrefs}).

In a sense then, by allowing Dirichlet boundary conditions, the string theory we developed in the previous chapter already contained room for open string T-duality; we didn't have to extend the formalism. But of course, in a very important way something very new has just been discovered; for what are the ends attached to? The reason we didn't pursue Dirichlet boundary conditions to start with is that they involve invariant locations, and so violate Poincar\'e invariance (and momentum conservation at the ends of the string). But there is another possibility. Suppose that a more complete theory, to which our perturbative string theory approximates, contains other dynamical (properly relativistic) objects than strings; specifically multi-dimensional generalizations of strings, known as $p$-branes. And suppose further that open strings can attach to them, enforcing Dirichlet boundary conditions: that they are `D$p$-branes'.\footnote{Where `D' is for `Dirichlet', and $p$ is the dimensionality of the brane. $0\geq p\geq D$ (the dimension of space); and, if you think about it, if the end of the string is attached to a brane, then it is fixed in $D-p$ dimensions, and free to move in $p$ of them. For instance, a point attached to the $z$-axis in 3-space can only move in one dimension, and is fixed in the $x$-$y$-plane.} And even further that they are non-perturbative: the approximations on which perturbative string theory is based, means that they do not appear as dynamical objects in the theory. If so, then open string T-duality simply shows that perturbative string theory is not completely independent of D$p$-branes, but recognizes their physical presence, at the level of non-dynamical objects.

In that case, one would very much like to know whether more can be discovered about D$p$-branes at the perturbative level, and of course this question has been extensively studied.\footnote{For a sense of their physical significance, refer to the references of footnote \ref{ftnt:ostdrefs}. For philosophical discussion of their nature see \cite{Vis:17} and \citet[chapter 4]{Vis:19}. } Those results fall outside the scope of this work, but it is worth noting that the gauge fields that arise in superstring theory naturally couple to `charges' carried by $p$-branes. In other words, their presence is revealed in perturbative string theory in more ways than through T-duality. Moreover, this role opens the door to a great deal of very interesting (and physically realistic) physics.

We promised above to apply these considerations to the question of whether T-dual theories could be realistically quotiented by a pair of wavefunctions. The presence of D$p$-branes as physical objects, makes this proposal even less realistic since they are not described in the two field approximation we used above. The pair of wavefunctions carry no information about branes or their locations; indeed, any quotient would have to be the same regardless of the presence or absence of branes, so insensitive to them in that sense. What one expects instead is that the duals are `quotiented' by an underlying theory, in which spacetime concepts do not fully apply; but which can equally be approximated with and without branes, or by any pair of dual theories. But at the level of perturbation theory, around a spacetime solution, we have to give significance to spatiotemporal concepts; that they are artifacts of an approximation scheme leads to their surprising indeterminacy.

\subsection{Gauge-gravity duality}\label{sec:GGD}

In the early 21st century, perhaps the most important tool for the study of string theory has been what is variously known as `AdS-CFT duality' (an acronym of `Anti de Sitter-conformal field theory'), or `gauge-gravity duality' (the gauge theory being the CFT, and the gravity theory AdS space), or `holographic duality' (though strictly this is a more general concept). This duality was proposed by Maldacena, whose argument we will briefly discuss. But it has its roots in an earlier proposal by 't Hooft \citep{SteHoo:94} based on the observation that the entropy of a black hole is proportional to its area not volume, suggesting that the entropy counts states on the `boundary' not `bulk'. In other words, suggesting that the full state of the black hole can be described by microstates on its horizon. Of course, for appropriate densities of states it's always possible to match the number of states on bulk and boundary; but the point is that for a constant  quantized state density the number of states will not remain equal -- doubling the radius means $4\ \times$ the boundary states but $8\ \times$ the bulk states. So the idea that bulk and boundary of any system can be physically equivalent -- dual -- is remarkable: but it seems to be the case when one compares a theory of gravity in AdS space, with a conformal gauge field on its boundary.

Maldacena's argument (\cite{Mal:98})\footnote{See \cite[chapter 23]{Zwi:04} or \cite{Daw:17} for intuitive presentations.} rests on the string theoretic understanding of gauge fields on the one hand, and of gravity and curved spacetime on the other. First then, gauge quanta can be understood in terms of the modes of open strings whose ends are attached to coincident D$p$-branes. Specifically, if there are $N$ such branes then the ground states of (supersymmetric) strings connecting them will describe the quanta of an $SU(N)$ Yang-Mills gauge field -- just as in the previous chapter we understood the quanta of bosonic fields as modes of bosonic strings when viewed `close up'.\footnote{More carefully, they are quanta of a $U(N)$ field, which decouples into $SU(N)$ plus a single gauge field.} Moreover, the branes and strings will carry charge for this gauge field. Second, we know that string modes have mass because of their tension, and the same applies to branes. Then the energy-momentum of the gauge field, its charges, and of the string and brane masses will provide a source for a relativistic gravitational field: the matter side of the Einstein field equation.\footnote{In the next chapter we will explicate the nature of gravity in string theory in detail. Ultimately the curved geometry is understood in terms of graviton states of the string, so that instead of sourcing an extrinsic gravitational field (as it seems here), matter and gravitational fields are both states of strings; the theory of gravity is a theory of string-string interaction.} 

In other words both the gauge field on the boundary and gravity in the bulk can both be understood in string terms if one focusses in to small enough scales; and AdS-CFT duality ultimately asserts the equivalence of apparently very different systems of strings. Maldacena's argumentative strategy for this duality was to compare a system at weak and strong  string couplings; observe that a low energy equivalence holds for half of it; and conjecture (on the basis of additional evidence) that the equivalence also holds for the other half: a gauge theory at weak coupling and AdS at strong. (Note that the string coupling is a dynamic quantity, not a constant of nature, and so really will vary.) 

In slightly more detail: the system in question consists of  closed superstrings, and $N$ coincident D3-branes with attached open superstrings, all in flat 10-dimensional spacetime. For weak coupling: there are no interactions between the strings, so the systems decouple; moreover, gravity is `turned off', and so spacetime is flat; and finally, at low energy the open strings will be unexcited, massless quanta of an $SU(N)$ gauge field. For strong coupling: gravity is now `turned on', and spacetime will be curved by the energy and charge of the D3-branes; near the branes the geometry will $AdS_5$ (from the 4 spacetime dimensions parallel to the branes plus the radial dimension) times $S^5$ from the remaining dimensions; far from the branes the geometry will be flat; then plausibly the closed strings near the branes do not interact with those far away, leaving (at low energy) a stringy gravitational field in $AdS_5\times S^5$ and again closed strings in flat spacetime. The coincidence of the closed string part of the system at weak and strong couplings then supports the idea that the other part of the system -- a gauge theory at weak coupling, and  $AdS_5\times S^5$ gravity at strong -- also coincides, so that these theories are equivalent, or rather \emph{dual}.

There has been considerable philosophical discussion of AdS-CFT\footnote{A good cross-section is: \cite{De-May:16,HarTeh:16,Har:17,Mat:13,Pol:17,Read:2014fk,Ric:12,Teh:2013rc,Vis:17,Vis:19}.}; we have treated  the general philosophical consequences of duality in terms of T-duality instead, and a discussion of the more specific features of AdS-CFT is largely beyond the scope of this work. However, we want to recommend an essay by \citet{Har:20} which presents a simple but illuminating and contentful model of AdS-CFT duality, in which the bulk contains a black hole. Briefly, three `qutrits'\footnote{A qutrit is the 3-dimensional generalization of an ordinary (2-dimensional) qubit.} live on the boundary, comprising a $3^3=27$-dimensional Hilbert space of a boundary quantum theory. The bulk theory is represented by a single qutrit, living in a 3-dimensional subspace of the full theory; corresponding to the few degrees of freedom of a classical black hole. But the bulk theory should be dual to that on the boundary, and so also live in a 27-dimensional Hilbert space; what has happened to the other 24 dimensions?  Harlow's point is that although both bulk and boundary theories are effective descriptions of more fundamental string states, the boundary \emph{quantum} gauge theory is still more fundamental -- closer to string theory -- than the bulk \emph{classical} theory. When one recognizes that point it is natural to understand the missing 24 dimensions as representing \emph{quantum} microstates of the black hole. If so, AdS-CFT duality allows one to study the unknown quantum nature of black holes through the better understood physics of CFT. Harlow's work explores and supports this very idea.

He emphasizes that this picture also clarifies the common view that AdS-CFT duality asserts the equivalence of the two theories. This view is correct at the level of the fundamental string description, but not at the level of the effective AdS and effective CFT descriptions, for the latter is a more complete description than the former; as we just saw, it carries more information about the fundamental string state. In that sense, the classical  AdS theory is not equivalent to the quantum  CFT, but \emph{derived} from it; then as we saw, the additional boundary degrees of freedom allow one to probe quantum gravity in the bulk. Moreover, this is a picture in which not all quantum degrees of freedom correspond to classical spacetime degrees of freedom, suggesting -- with Brandenberger and Vafa -- that the fundamental ones may not be spatiotemporal at all.\footnote{See \cite{Mat:12} for an introduction to the `fuzzball' approach to string theoretic black holes, which proposes a specific quantum structure  (and \citet{Nic:20a} for further philosophical analysis). In a fuzzball model spacetime indeed `ends' at the horizon, leaving the black hole state `beyond spacetime'.}\\

\section{Conclusions}

The main conclusions of this chapter are as follows. First, T-duality is an unusually deep symmetry between theories, with respect to some very counterintuitive and surprising parameters: especially the radius of space. Gauge symmetries in field theory are similarly deep, but since they typically involve internal degrees of freedom, they are not so shocking. A touchstone of this chapter has been the analysis of Brandenberger and Vafa, which explains how there can be two theories  apparently differing on the radius of space, yet predicting the same observed radius. Their analysis has helped at several points to understand the physical meaning of T-duality: such a picture is crucial to understanding duality.

The symmetry is so deep -- between all observables, not just empirical quantities in some superficial sense -- that duals should be understood as giving physically equivalent descriptions. Since they formally disagree on some claims, we have argued (against an alternative view) that the physical commitments of dual theories are limited to their common consequences. Specifically, they disagree on the radius of target space, so that must be indeterminate between the two possible values. And in general, `target space' is not a space in the familiar sense at all, but a `space' with only the structures on which the duals agree. (Quite possibly then, a structure that appears as a formal representation of some more fundamental, as yet unknown, non-spatial object.) As the analysis of Brandenberger and Vafa explains, duals do agree on the radius of phenomenal space, so that is determinate. But nothing can be both determinate and indeterminate with respect to radius, and so target space is not classical, relativistic space. 

Therefore classical space, specifically as a geometric space of determinate radius, is not a fundamental object of string theory, but an appearance, arising from physical processes of the kind that Brandenberger and Vafa analyzed. That, ultimately, is the ontological significance of T-duality, and indeed of the other dualities we have described.\\

\bibliographystyle{plainnat}
\bibliography{../../Bibliography/biblio}

\end{document}